\documentclass[manuscript,screen]{acmart}

\usepackage{hyperref}
\usepackage{url}
\usepackage{graphicx}
\usepackage[table,xcdraw]{xcolor}
\usepackage{booktabs}
\usepackage{amsmath}
\usepackage{wrapfig} 
\usepackage{multirow}
\usepackage{setspace}
\usepackage{tabularx}
\usepackage{array}
\usepackage{makecell}
\usepackage{caption}
\usepackage{adjustbox}
\usepackage{enumitem}
\usepackage{algorithm}
\usepackage{algorithmic}
\usepackage{multirow} 
\usepackage{enumitem}
\usepackage{makecell}
\usepackage{algorithm}
\usepackage[utf8]{inputenc}
\usepackage{amsmath}
\usepackage{amsfonts}
\usepackage{algorithm}
\usepackage{algorithmic}
\usepackage{booktabs}
\usepackage{multirow}
\usepackage{graphicx}
\usepackage[normalem]{ulem}
\useunder{\uline}{\ul}{}

\definecolor{lightblue}{rgb}{0.85,0.89,0.96}
\definecolor{lightgray}{rgb}{0.95,0.95,0.95}
\definecolor{accentblue}{rgb}{0.2,0.4,0.7}
\definecolor{lightyellow}{rgb}{1,0.98,0.9} 
\AtBeginDocument{%
  }

\setcopyright{acmlicensed}
\copyrightyear{2018}
\acmYear{2018}
\acmDOI{XXXXXXX.XXXXXXX}
\acmConference[Conference acronym 'XX]{Make sure to enter the correct
  conference title from your rights confirmation email}{June 03--05,
  2018}{Woodstock, NY}
\acmISBN{978-1-4503-XXXX-X/2018/06}

\begin{document}

\title{Looking Farther with Confidence: Uncertainty-Guided Future Learning for Sequential Recommendation}

\author{Ziqiang Cui}
\email{ziqiang.cui@my.cityu.edu.hk}
\affiliation{%
  \institution{City University of Hong Kong}
  \city{Hong Kong}
  \country{China}
}

\author{Xing Tang}
\email{xing.tang@hotmail.com}
\affiliation{%
  \institution{Shenzhen Technology University}
  \city{Shenzhen}
  \country{China}
}

\author{Peiyang Liu}
\email{liupeiyang@pku.edu.cn}
\affiliation{%
  \institution{Peking University}
  \city{Beijing}
  \country{China}
}

\author{Xiaokun Zhang}
\email{dawnkun1993@gmail.com}
\affiliation{%
  \institution{City University of Hong Kong}
  \city{Hong Kong}
  \country{China}
}

\author{Shiwei Li}
\email{lishiwei@hust.edu.cn}
\affiliation{%
  \institution{Huazhong University of Science and Technology}
  \city{Wuhan}
  \country{China}
}

\author{Xiuqiang He}
\email{hexiuqiang@sztu.edu.cn}
\affiliation{%
  \institution{Shenzhen Technology University}
  \city{Shenzhen}
  \country{China}
}

\author{Chen Ma}
\email{chenma@cityu.edu.hk}
\affiliation{%
  \institution{City University of Hong Kong}
  \city{Hong Kong}
  \country{China}
}

\renewcommand{\shortauthors}{Ziqiang Cui et al.}

\begin{abstract}
Sequential recommendation effectively models dynamic user interests but continues to face challenges related to data sparsity. While self-supervised learning has alleviated this issue to some extent, most existing methods focus exclusively on immediate next-item prediction during training, thereby neglecting the rich information embedded in longer-term future interactions. Although a few studies have explored the utilization of future data, existing attempts typically apply future supervision signals with uniform intensity across all samples, which may lead to suboptimal solutions. In this paper, we propose an adaptive future learning framework, \textbf{UFRec}, which encourages the model to look further ahead when it is confident in the current state, while focusing on the immediate task when it is uncertain. Specifically, UFRec incorporates an Uncertainty-Guided Future Supervision module that dynamically modulates the weight of multi-step future supervision based on the model's confidence in the primary next-item prediction task. Furthermore, we complement step-wise future supervision with a Future-Aware Contrastive Learning module that treats the future trajectory as a holistic entity. Notably, both auxiliary modules are utilized exclusively during training and incur no inference overhead. Extensive experiments on four benchmark datasets demonstrate that our method significantly outperforms state-of-the-art approaches by effectively leveraging future data. Our code is available at
\textcolor{blue}{\url{https://github.com/ziqiangcui/UFRec}}.
\end{abstract}




\maketitle

\section{Introduction}
Recommender systems play a critical role in alleviating information overload by filtering vast amounts of content and delivering personalized suggestions. To capture dynamic user interests, sequential recommendation has emerged as a key research area. Unlike static models that treat user interactions as isolated events, sequential recommendation explicitly models the temporal dependencies and evolving patterns within a user's historical behaviors, achieving notable results. Consequently, this paradigm has been widely adopted by major online platforms such as YouTube and Amazon.

Despite significant advancements in sequential recommendation, the field remains constrained by the sparsity of user-item interactions, a problem characterized by users with limited history and items receiving scant attention. To mitigate this challenge, numerous studies have introduced auxiliary self-supervised signals to enhance data utilization~\cite{zhang2024recdcl, wu2021self,xia2023automated, zhou2023selfcf,xu2023multi,qin2024intent,zhao2025survey}. This paradigm typically involves training on auxiliary objectives where ground-truth samples are derived directly from the raw data. A prominent example is S3-Rec \cite{zhou2020s3}, which employs four self-supervised objectives to learn correlations among attributes, items, subsequences, and sequences via mutual information maximization. Furthermore, contrastive learning has emerged as a dominant self-supervised approach by constructing positive pairs and minimizing their distance in the representation space to capture intrinsic patterns. Within this framework, some methods generate contrastive pairs by creating semantically related views of the same sequence via data augmentation \cite{xie2022contrastive,qiu2022contrastive,qin2023meta}, while others identify similar users as positive pairs through clustering or retrieval mechanisms \cite{chen2022intent,cui2025semantic,qin2024intent}.

\begin{figure*}
    \centering
    \includegraphics[width=0.8\textwidth]{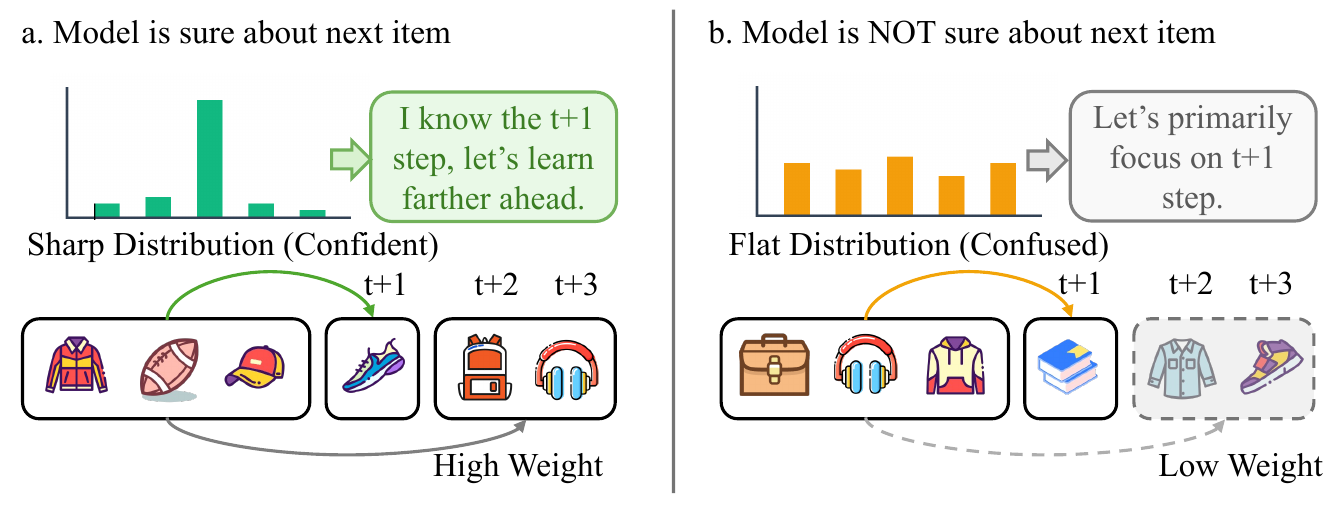}
    \caption{An illustration of our proposed Uncertainty-Guided Future Supervision.}
    \label{fig:intro}
\end{figure*}


While existing self-supervised methods have proven effective, they suffer from suboptimal data utilization due to their exclusive reliance on the immediate next item for supervision. This narrow focus discards the valuable information contained in longer-term interactions, leading to model myopia: representations struggle to capture evolving long-term preferences and become susceptible to short-term noise or accidental clicks. To address this, we propose leveraging multi-step future behaviors as auxiliary supervisory signals during training. While the primary task remains next-item prediction, incorporating these future interactions offers three distinct benefits. First, it transforms sparse one-step signals into denser supervision. Second, it encourages the model to look beyond the immediate horizon, thereby improving its understanding of interest evolution. Finally, by integrating behaviors over multiple steps, the model can more effectively filter out isolated anomalies to identify consistent user preferences, ensuring more robust supervision.

However, the effective utilization of future interactions remains a significant challenge. To the best of our knowledge, only two works have investigated this area: DSSRec~\cite{ma2020disentangled}, which predicts future intentions via clustering and disentangled representations, and FENRec~\cite{huang2025future}, which incorporates future supervision using time-dependent soft labels.
Crucially, both methods apply future supervision indiscriminately across all samples. Although predicting multi-step future behaviors can theoretically increase supervision density, enforcing such objectives uniformly is problematic. Specifically, if a model exhibits high uncertainty in predicting the immediate next item, it implies an inadequate understanding of the user's current preference. Compelling the model to predict distant interactions under these conditions introduces unreliable signals that may conflict with the primary task. This risks degrading the learned representations, thereby further impairing performance on the main objective (i.e., next item prediction). Therefore, we argue that the effective utilization of future data should take into account the model's confidence in the current state.

Motivated by this, we propose \textbf{U}ncertainty-Guided \textbf{F}uture
Learning for Sequential \textbf{Rec}ommendation (\textbf{UFRec}). UFRec is a model-agnostic framework designed to effectively leverage future interactions as auxiliary signals during training. 
UFRec incorporates an uncertainty-guided future supervision module. Specifically, we quantify the model's uncertainty in the primary next-item prediction task based on Shannon entropy, where the uncertainty is subsequently utilized to dynamically adjust the weight of the auxiliary future supervision loss. This mechanism ensures that the model assigns less weight to future supervision when it lacks confidence in the immediate next step. Conversely, when the model is confident in its current prediction, it is encouraged to leverage future information more aggressively. To further enhance representation quality, we complement the step-wise supervision with a future-aware contrastive learning module that treats the future trajectory as a holistic entity. This module explicitly aligns the user's current representation with their own ``future horizon'' while distinguishing it from the future trajectories of other users, thereby refining the learned representations.
Crucially, all auxiliary objectives in our framework serve strictly as training-time regularizations. During inference, the auxiliary modules are detached, incurring zero additional computational overhead compared to the backbone model. Extensive experiments on four benchmark datasets validate the superiority of our method.

Our contributions are summarized as follows:
\begin{itemize}
\item We propose an \textbf{Uncertainty-Guided Future Supervision} mechanism that adaptively leverages multi-step future interactions as auxiliary signals during training. By dynamically modulating the future supervision weight based on the uncertainty of the primary next-item prediction, we effectively utilize valuable future data while preventing negative interference with the primary task. 
\item We introduce a \textbf{Future-Aware Contrastive Learning} module to capture holistic preference trends beyond step-wise accuracy. This component aligns the user's current representation with their specific future horizon while distinguishing it from others, thereby capturing holistic preference evolution and enhancing the discriminative power of the learned representations.
\item We conduct extensive experiments on four real-world benchmark datasets. The results demonstrate that our framework significantly outperforms state-of-the-art sequential recommendation methods, and further analyses validate the effectiveness of each module and the framework's strong generalizability across different recommendation backbones.
\end{itemize}

\section{Related Work}
In this section, we review the literature closely related to our proposed framework. We first summarize the evolution of sequential recommendation models, highlighting the shift from traditional probabilistic methods to advanced deep learning-based architectures. Subsequently, we discuss the application of self-supervised learning in recommender systems. 
\subsection{Sequential Recommendation}
The progression of sequential recommendation (SR) methodologies has been marked by a shift from probabilistic modeling to advanced representation learning. Initially, user behavior sequences were typically formulated as Markov chains~\cite{rendle2010factorizing, he2016fusing}. However, the emergence of deep learning introduced powerful architectures like Convolutional Neural Networks (CNNs) and Recurrent Neural Networks (RNNs) to the domain, resulting in superior predictive capabilities~\cite{tang2018personalized,hidasi2015session,hidasi2018recurrent}. A major paradigm shift occurred with the adoption of attention mechanisms \cite{li2026r2ns,zhang2024disentangling}; SASRec~\cite{kang2018self} was the first to leverage self-attention for SR, a concept further extended by BERT4Rec~\cite{sun2019bert4rec} through bidirectional encoding to capture comprehensive context. Contemporary research continues to build on these foundations, with recent studies integrating graph neural networks and enhanced self-attention layers to further optimize recommendation performance~\cite{he2020lightgcn,zhang2022dynamic,xia2022multi,su2025intrinsic}.
Recently, the field has witnessed a surge of interest in adapting Large Language Models (LLMs) for SR \cite{wang2025generative,lin2025can}. Generally, LLMs are employed either as direct recommenders or as enhancers for extracting semantic information~\cite{li2023prompt,liao2024llara,zhao2024let,boz2024improving,ren2024enhancing}. In the former paradigm, user behaviors and item descriptions are converted into textual prompts, allowing the LLM to generate recommendations directly based on its pre-trained knowledge or through supervised fine-tuning. In the latter paradigm~\cite{ren2024enhancing,wang2024can,liu2024llm,yang2024sequential,zhao2024let,boz2024improving}, LLMs are leveraged to process textual data and integrate rich semantic representations into traditional ID-based models, thereby combining the reasoning power of LLMs with the efficiency of collaborative filtering. However, all these methods rely on the immediate next item as the supervision label, thereby overlooking the valuable information embedded in future items.

\subsection{Self-Supervised Learning}
Self-supervised learning trains networks using auxiliary objectives, where ground-truth labels are automatically derived from raw data \cite{yu2023self,ren2025comprehensive}. It has been widely adopted to mitigate data sparsity by constructing rich supervisory signals directly from the data itself~\cite{yang2022knowledge,zhang2024recdcl, yu2022graph, wu2021self,xia2023automated,wang2023poisoning, zhou2023selfcf,xu2023multi,jiang2024diffkg,yin2022autogcl,cui2024context,qin2024intent,jaiswal2020survey,zhao2025survey,xia2025oracle}. For instance, $\rm {S^3\text{-}Rec}$\cite{zhou2020s3} devises four auxiliary self-supervised objectives to learn correlations among attributes, items, subsequences, and sequences via mutual information maximization. In addition, BERT4Rec\cite{sun2019bert4rec} adopts a Cloze objective, predicting randomly masked items within a sequence by conditioning jointly on their left and right contexts. However, most existing methods fail to leverage future interactions. To address this, \citet{ma2020disentangled} proposed a disentangled sequence-to-sequence self-supervision strategy that mines long-term signals by reconstructing future sequence representations in the latent space. Furthermore, \citet{huang2025future} introduced FENRec, which enhances sequential recommendation by employing time-dependent soft labeling to utilize future interactions and incorporating enduring hard negative samples. While these two works demonstrate the value of future signals, they suffer from a common drawback: they apply future supervision indiscriminately to all samples. When the model exhibits high uncertainty in predicting the immediate next item—indicating an insufficient grasp of the user's current preference—forcing it to predict distant future interactions introduces unreliable supervisory signals. This noise propagation can degrade the quality of learned representations. In contrast to these static approaches, our method introduces an uncertainty-aware mechanism. We quantify the model's uncertainty in the primary next-item prediction task and dynamically adjusts the weight of the auxiliary future supervision loss. This ensures that the model assigns lower importance to future supervision when confidence in the immediate next step is low, thereby adaptively suppressing unreliable signals derived from uncertain states.

As a pivotal paradigm in self-supervised learning, contrastive learning derives informative supervisory signals from unlabeled data and has been widely adopted in recent years. For instance, CL4SRec~\cite{xie2022contrastive} employs three data-level augmentation operators (cropping, masking, and reordering) to generate positive pairs, thereby promoting representational invariance. In addition, CoSeRec \cite{liu2021contrastive} proposes substituting specific items within a sequence with semantically similar alternatives. At the model level, DuoRec~\cite{qiu2022contrastive} generates positive pairs by forward-passing an input sequence twice using distinct dropout masks. Additionally, ICLRec~\cite{chen2022intent} extracts user intent from sequential data to perform contrastive learning between user and intent representations. More recently, SRA-CL~\cite{cui2025semantic} leverages the semantic understanding capabilities of LLMs to guide the construction of contrastive samples, generating more rational positive pairs and further improving performance. Despite the effectiveness of these approaches, they fail to incorporate future interactions within user sequences. In contrast to existing methods, we propose a future-aware contrastive learning method, which aligns the user’s current representation with a pooled representation of their ``future horizo'' while distinguishing it from the future trajectories of other users.

\section{Problem Formulation}
In this work, we focus on the task of sequential recommendation, which aims to predict the next item a user will interact with by modeling their historical interaction patterns. Let $\mathcal{U}$ and $\mathcal{V}$ denote the sets of users and items, respectively. For each user $u \in \mathcal{U}$, the historical interactions are arranged chronologically as a sequence $\mathcal{S}_u = [v^u_1, v^u_2, \dots, v^u_{|\mathcal{S}_u|}]$, where $v^u_i \in \mathcal{V}$ represents the item interacted with at the $i$-th step, and $|\mathcal{S}_u|$ denotes the total number of interactions for user $u$.
Given a user interaction sequence $\mathcal{S}_u$, the primary objective is to predict the item with which the user is most likely to interact at the next time step.

During the training phase, existing methods often divide the original sequence into multiple subsequences to enrich the training data \cite{kang2018self,qiu2022contrastive,qin2023meta,huang2025future}. Specifically, each user sequence $\mathcal{S}_u$ is processed into multiple training instances. A single training instance is defined by a prefix subsequence $\mathcal{S}^u_{1:t} = [v^u_1, v^u_2, \dots, v^u_t]$, where $1 < t+1 \le |\mathcal{S}_u|$. Formally, the entire training set $\mathcal{D}$ is constructed by collecting all valid input-target pairs across all users:
\begin{equation}
    \mathcal{D} = \{ (\mathcal{S}^u_{1:t}, v^u_{t+1}) \mid u \in \mathcal{U}, 1 < t+1 \le |\mathcal{S}_u| \}.
\end{equation}
The objective for each instance is to predict the immediate next item $v^u_{t+1}$ given the context $\mathcal{S}^u_{1:t}$. Consequently, the model is trained to minimize the negative log-likelihood:
\begin{equation}
    \mathcal{L} = - \sum_{u \in \mathcal{U}} \sum_{t=1}^{|\mathcal{S}_u|-1} \log P(v^u_{t+1} \mid \mathcal{S}^u_{1:t}).
\end{equation}

In this paper, while we adhere to the standard sequential recommendation formulation, we extend the training paradigm by introducing future interactions beyond the immediate next item (i.e., \(v^u_{t+2}, \dots, v^u_{t+K}\)) as auxiliary supervision signals. It should be noted that these future signals are exclusively utilized during the training phase to enhance representation learning. During the inference phase, our framework strictly follows the standard setting, focusing solely on predicting the immediate next item \(v^u_{t+1}\) based on the historical context \(\mathcal{S}^u_{1:t}\).

\begin{figure*}[t]
  \centering
  \includegraphics[width=\textwidth]{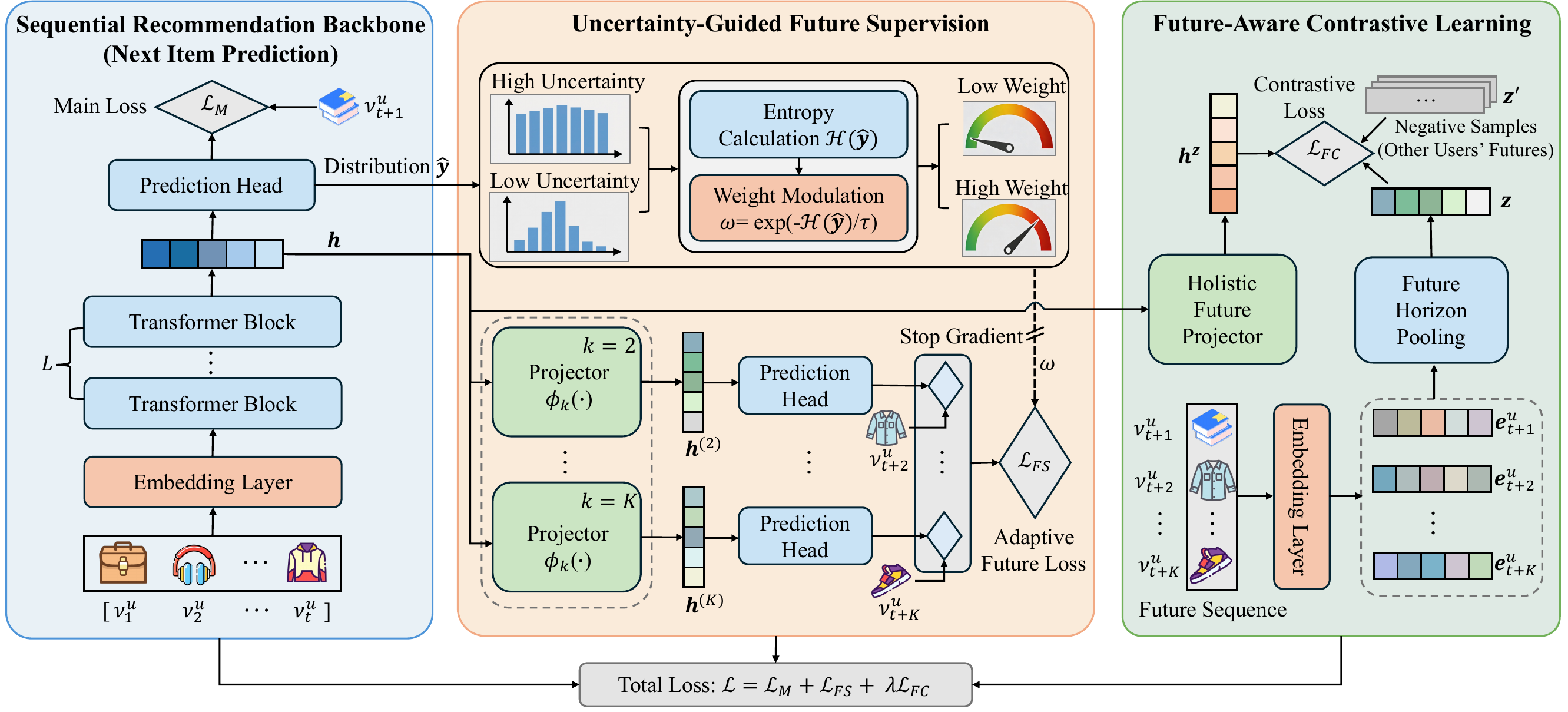}
\caption{Overview of the proposed UFRec framework. (1) The {sequential recommendation backbone} (left) encodes the user's historical interaction sequence into a latent state $\mathbf{h}$ to predict the immediate next item $v_{t+1}^u$. (2) The {uncertainty-guided future supervision} module (middle) performs parallel multi-step future prediction. It incorporates a novel uncertainty-aware mechanism that computes the entropy of the main prediction to dynamically modulate the auxiliary loss weight $\omega$. (3) The {future-aware contrastive learning} module (right) captures global preference trends by aligning the projected current state $\mathbf{h}^z$ with a future horizon representation $\mathbf{z}$ derived from the ground-truth future sequence. Note that components (2) and (3) are only utilized during training and introduce no additional computational overhead during inference.}
\label{fig:method}
\end{figure*}

\section{Methodology}

\begin{table}[htbp]
  \centering
  \caption{Summary of notations used in the UFRec model}
  \label{tab:notations}
  \renewcommand{\arraystretch}{1.2} 
  \begin{tabular}{cl}
    \toprule
    \textbf{Notation} & \textbf{Description} \\
    \midrule
    \(\mathcal{U}, \mathcal{V}\) & Sets of users and items, respectively. \\
    \(\mathcal{S}_u\) & The historical interaction sequence of user \(u\). \\
    \(v^u_t\) & The item interacted with by user \(u\) at the \(t\)-th step. \\
    \(\mathbf{M}\) & The item embedding matrix. \\
    \(\mathbf{h}\) & The final representation of the user's current state at step \(t\). \\
    \(\hat{\mathbf{y}}\) & The predicted probability distribution over candidate items for the next step. \\
    \(K\) & The number of future steps for auxiliary supervision. \\
    \(\mathbf{h}^{(k)}\) & The user intent projected \(k\) steps into the future. \\
    \(\omega\) & The uncertainty-guided adaptive weight for future supervision. \\
    \(\mathbf{z}\) & The future horizon representation derived from ground-truth future items. \\
    \(\mathbf{h}^z\) & The projected user preference representation for contrastive learning. \\
    \(\mathcal{L}_{M}\) & Main objective loss for next-item prediction. \\
    \(\mathcal{L}_{FS}\) & Uncertainty-guided future supervision loss. \\
    \(\mathcal{L}_{FC}\) & Future-aware contrastive learning loss. \\
    \bottomrule
  \end{tabular}
\end{table}

This section details the proposed UFRec framework, as illustrated in Figure \ref{fig:method}. We first formulate the sequential recommendation backbone, then introduce the uncertainty-guided future supervision and future-aware contrastive learning modules. Finally, we introduce the training objective of our approach. For clarity, Table \ref{tab:notations} summarizes the key notations used throughout this paper.

\subsection{Sequential Recommendation Backbone} \label{sec:backbone}

Our proposed framework is inherently model-agnostic and designed for seamless integration with various sequential recommendation models. To facilitate the presentation of our approach and in accordance with established research protocols \cite{qin2023meta,qin2024intent,qiu2022contrastive}, we employ the Transformer architecture \cite{vaswani2017attention} as the representative backbone model.

\subsubsection{\textbf{Embedding Layer}}  \label{emb layer}
We utilize an item embedding matrix $\mathbf{M} \in \mathbb{R}^{|\mathcal{V}| \times d}$ to project item IDs into a $d$-dimensional latent space, where $|\mathcal{V}|$ denotes the size of the item vocabulary. Given a sequence $\mathcal{S}^u_{1:t} = [v^u_1, v^u_2, \dots, v^u_t]$, we retrieve the corresponding item embeddings from the embedding matrix to form $\mathbf{E} \in \mathbb{R}^{t \times d}$. To preserve position information, we add learnable position embeddings $\mathbf{P}\in \mathbb{R}^{t \times d}$. The final input representation of the sequence is formulated as $\mathbf{H} = \mathbf{E} + \mathbf{P}$, where $\mathbf{H} \in \mathbb{R}^{t \times d}$.

\subsubsection{\textbf{Sequence Encoder}} \label{user seq enc}
Following the embedding layer, the input vector sequence $\mathbf{H}$ propagates through $L$ Transformer blocks \cite{vaswani2017attention} to learn complex sequential patterns. Each Transformer block consists of a self-attention layer and a feed-forward network. This process is formulated as:
\begin{equation} \label{rec_sequence_representation}
\begin{split}
    \mathbf{H}^{L} &= \text{Transformer}(\mathbf{H}), \\
    \mathbf{h} &= \mathbf{H}^{L}[t],
\end{split}
\end{equation}
where $\mathbf{H}^{L}$ denotes the output of the final layer. We utilize the vector at the last position $t$, denoted as $\mathbf{h} \in \mathbb{R}^{d}$, as the final representation of the user's current state.

\subsubsection{\textbf{Main Objective}} 
Based on the user representation at time step $t$, we calculate the probability distribution over all candidate items. It is computed as:
\begin{equation} \label{eq:y}
    \hat{\mathbf{y}} = \operatorname{softmax}(\mathbf{h} \mathbf{M}^\mathrm{T}),
\end{equation}
where $\hat{\mathbf{y}} \in \mathbb{R}^{|\mathcal{V}|}$, with $\hat{y}_v$ representing the probability of item $v$ being the next interaction. 
For training, we adopt the same cross-entropy loss function as our baseline methods \cite{qiu2022contrastive,qin2024intent,huang2025future}:
\begin{equation} \label{rec_loss}
\mathcal{L}_{{M}} = -\log \left( \hat{y}_{v^u_{t+1}} \right),
\end{equation}
where $\hat{y}_{v^u_{t+1}}$ specifically denotes the predicted probability assigned to the ground truth next item $v^u_{t+1}$.

\subsection{Uncertainty-Guided Future Supervision}

While the primary objective is to predict the immediate next item $v^u_{t+1}$, relying solely on this single signal may lead to short-sightedness and sparse data supervision, causing suboptimal representation learning. To address this, we propose leveraging future interactions beyond the next one as \textit{auxiliary supervision signals}. 
It is important to note that this auxiliary task is conducted only during the training phase. Since the training sample $\mathcal{S}^u_{1:t} = [v^u_1, v^u_2, \dots, v^u_t]$ is a subsequence extracted from the user's complete history $\mathcal{S}_u$, we have access to the subsequent ground-truth items in $\mathcal{S}_u$ beyond time step $t$, provided the sequence has not ended. 


\subsubsection{\textbf{Parallel Multi-Step Projection}} 
While standard models typically predict only the immediate next step ($t+1$), we extend our model to predict $K-1$ additional consecutive future steps ($t+2, \dots, t+K$). Unlike autoregressive approaches that generate predictions recursively, our method performs these additional $K-1$ predictions simultaneously in a parallel manner, ensuring efficient generation.

Let $\mathbf{h}$ denote the sequence representation at step $t$ used for the next-item prediction task (as defined in Equation \ref{rec_sequence_representation}). To avoid interfering with the latent space of the main task, we employ separate projection heads. Specifically, for the $k$-th future step ($1 < k \le K$), we transform the current state $\mathbf{h}$ via a step-specific projector $\phi_k(\cdot)$:
\begin{equation} 
\label{eq:projector}
    \mathbf{h}^{(k)} = \phi_k(\mathbf{h}),
\end{equation}
where $\mathbf{h}^{(k)}$ represents the user intent projected $k$ steps into the future from time $t$ and $\mathbf{h}^{(1)} = \mathbf{h}$. Each projector is designed to be lightweight, comprising a single linear layer followed by a ReLU activation. Crucially, to ensure high training efficiency, these $K-1$ projections are computed in parallel from the base representation $\mathbf{h}$.
Given these projected representations, we can compute the probability distribution over items for the $k$-th future step (i.e., $(t+k)$-th time step), denoted as $\hat{\mathbf{y}}^{(k)}$, as follows:
\begin{equation}
    \hat{\mathbf{y}}^{(k)} = \operatorname{softmax}(\mathbf{h}^{(k)} \mathbf{M}^\mathrm{T}).
\end{equation}

\subsubsection{\textbf{Uncertainty-Guided Modulation}} 
However, indiscriminately incorporating future predictions can be detrimental if the current state estimation is unreliable. Since the primary objective of sequential recommendation is next item prediction, incorporating distant future supervision signals when the immediate next item prediction is inaccurate can introduce noise and exacerbate model performance issues. To resolve this, we introduce an uncertainty-aware mechanism that dynamically modulates the intensity of future supervision.
Specifically, we first quantify the uncertainty of the main task using the Shannon Entropy \cite{shannon1948mathematical}:
\begin{equation}
    \mathcal{H}(\hat{\mathbf{y}}) = - \sum_{v \in \mathcal{V}} \hat{y}_v \cdot \log \left(\hat{y}_v\right).
\end{equation}
Here, $\hat{\mathbf{y}}$ denotes the model's predicted probability distribution for the next item, as defined in Equation \ref{eq:y}. A high entropy value implies high uncertainty, suggesting that the model struggles to perform the primary task of next-item prediction; conversely, a low entropy value indicates that the model is confident in its current prediction.

Based on the entropy, we design a weighting function to adaptively modulate the intensity of future supervision signals in response to changes in model uncertainty. Specifically, the weight is formulated as:
\begin{equation} \label{eq:weight}
    \omega = \exp\left(-\frac{\mathcal{H}(\hat{\mathbf{y}})}{\tau}\right),
\end{equation}
where the temperature parameter $\tau$ governs the sensitivity of the decay, determining how strictly high-entropy states are penalized. The exponential function maps non-negative entropy values to a normalized range of $(0, 1]$. Consequently, when the model is highly confident (low entropy), the weight of the future supervision loss approaches $1$, encouraging the model to learn farther ahead. Conversely, as uncertainty increases, the weight decays toward $0$; this assigns minimal importance to distant future supervision, forcing the model to prioritize the current main task prediction. Notably, $\omega$ is detached from the computation graph to prevent gradient backpropagation.

\subsubsection{\textbf{Future Supervision Objective}} 
To formulate the future supervision objective, we employ the cross-entropy loss to minimize the discrepancy between the predicted distribution $\hat{\mathbf{y}}^{(k)}$ and the ground truth $v^u_{t+k}$ for each future step $k \in {2, \dots, K}$. This loss is computed for all $K-1$ future steps but is crucially modulated by an uncertainty-guided weight specific to the current sample to control its influence dynamically. The final learning objective is defined as:
\begin{equation}
    \mathcal{L}_{{FS}} = - \omega \cdot \frac{1}{K-1} \sum_{k=2}^{K} \log \left( \hat{y}^{(k)}_{v^u_{t+k}} \right),
\end{equation}
where $\omega$ modulates the magnitude of the future loss, as defined in Equation \ref{eq:weight}. Unlike a static hyperparameter, $\omega(\hat{\mathbf{y}})$ is an instance-dependent variable. When the model is confident in predicting the next step for the current sample, it assigns a larger weight to more distant steps; conversely, when the model encounters an uncertain sample, it reduces the focus on future steps. This mechanism ensures that the auxiliary future supervision is dynamically weighted according to the model's confidence in the next-item prediction for the current sample.
Let $|\mathcal{S}_u|$ denote the length of the user's full interaction sequence. Since the farthest future target item $v^u_{t+K}$ is available only if the user's history is sufficiently long, this future supervision loss is computed only for samples where $t+K \le |\mathcal{S}_u|$.

\subsection{Future-Aware Contrastive Learning}

While the uncertainty-guided future supervision task focuses on step-wise accuracy, we further introduce a future-aware contrastive learning task which treats the future trajectory as a whole. The core intuition is that the latent representation of a user's current history should be more semantically similar to their own future trajectory than to the future trajectories of other users.

\subsubsection{\textbf{Future Horizon Pooling}}
To represent the holistic future preference trend rather than individual step-wise item IDs, we construct a ``Future Horizon'' representation using a pooling operation. For a training sequence $\mathcal{S}^u_{1:t}$, given the ground-truth subsequent sequence $[v^u_{t+1}, \dots, v^u_{t+K}]$, we aggregate the corresponding item embeddings via mean pooling:
\begin{equation}
    \mathbf{z} = \frac{1}{K} \sum_{k=1}^{K} \mathbf{e}^u_{t+k},
\end{equation}
where $\mathbf{e}^u_{t+k}$ denotes the embedding of the item at step $t+k$. The vector $\mathbf{z}$ serves as a condensed anchor of the user's future interests. Note that this calculation is performed only for samples that possess a complete subsequent sequence of $K$ future interactions in the training set.

\subsubsection{\textbf{Holistic Future Projection}}
Directly aligning the user preference representation $\mathbf{h}$ with the future horizon representation $\mathbf{z}$ may conflict with the main task, as $\mathbf{h}$ is primarily optimized for next-item prediction. To address this, we project $\mathbf{h}$ into a separate latent space, which facilitates the contrastive learning task without interfering with the main objective. Formally, the holistic future projection is defined as:
\begin{equation}
\label{eq:future_project}
\mathbf{h}^{z} = \mathbf{W} \mathbf{h} + \mathbf{b},
\end{equation}
where $\mathbf{W} \in \mathbb{R}^{d \times d}$ and $\mathbf{b} \in \mathbb{R}^{d}$ are learnable parameters.

\subsubsection{\textbf{Future Contrastive Objective}}
We employ the InfoNCE loss to perform the contrastive learning task. In our contrastive framework, the user's future horizon representation $\mathbf{z}$ is regarded as the positive sample for the projected user preference representation $\mathbf{h}^z$, while the remaining $N-1$ future horizon representations ${\mathbf{z}^\prime \in \mathbf{Z}^\prime}$ of other users within the same batch are treated as negative samples, where $N$ is the batch size. The objective is to maximize the similarity of the positive pair while minimizing the similarity with negative pairs:
\begin{equation} \label{eq:loss_cl}
\mathcal{L}_{FC} = -\log \frac{\exp(\operatorname{sim}({\mathbf{h}^z}\cdot \mathbf{z}))}
{{\exp(\operatorname{sim}({\mathbf{h}^z}\cdot \mathbf{z}))}+ \sum_{\mathbf{z}^\prime \in \mathbf{Z}^\prime}{\exp(\operatorname{sim}({\mathbf{h}^z}\cdot \mathbf{z}^\prime))} },
\end{equation}
where $\operatorname{sim}(\cdot, \cdot)$ is the dot product similarity, $\mathbf{Z}^\prime$ denotes the set of negative samples for $\mathbf{h}^z$ in the same batch. Optimizing this objective enables the model to discriminate the current user's future horizon from those of others by leveraging the projected user preference representation, thereby improving representation quality.

\begin{algorithm}[t!]
\caption{Training Procedure of the Proposed Framework}
\label{alg:training}
\renewcommand{\baselinestretch}{1.2}\selectfont 
\begin{algorithmic}[1]
\REQUIRE Training set \(\mathcal{D}\), Hyperparameters \(K, \tau, \lambda\).
\ENSURE Trained model parameters \(\Theta\).
\STATE Initialize model parameters \(\Theta\) (including item embedding matrix \(\mathbf{M}\)).
\FOR{each iteration}
    \STATE Sample a sequence prefix \(\mathcal{S}^u_{1:t}\) from \(\mathcal{D}\).
    
    \STATE \textbf{// 1. Sequential Recommendation Backbone}
    \STATE Compute input embeddings \(\mathbf{H}\) corresponding to \(\mathcal{S}^u_{1:t}\).
    \STATE Encode sequence: \(\mathbf{H}^{L} = \text{Transformer}(\mathbf{H})\).
    \STATE Obtain current state representation: \(\mathbf{h} = \mathbf{H}^{L}[t]\).
    \STATE Compute main prediction: \(\hat{\mathbf{y}} = \operatorname{softmax}(\mathbf{h} \mathbf{M}^\mathrm{T})\).
    \STATE Calculate main task loss: \(\mathcal{L}_{{M}} = -\log(\hat{y}_{v^u_{t+1}})\).
    
    \STATE \textbf{// 2. Uncertainty-Guided Future Supervision}
    \STATE Calculate entropy: \(\mathcal{H}(\hat{\mathbf{y}}) = - \sum \hat{y}_v \log(\hat{y}_v)\).
    \STATE Compute adaptive weight: \(\omega(\hat{\mathbf{y}}) = \exp(-\mathcal{H}(\hat{\mathbf{y}}) / \tau)\) (detached).
    \STATE Project to future steps in parallel: \(\{\mathbf{h}^{(k)}\}_{k=2}^K = \{\phi_k(\mathbf{h})\}_{k=2}^K\).
    \STATE Compute weighted future loss: \(\mathcal{L}_{{FS}} = - \frac{\omega(\hat{\mathbf{y}})}{K-1} \sum_{k=2}^{K} \log(\hat{y}^{(k)}_{v^u_{t+k}})\).
    
    \STATE \textbf{// 3. Future-Aware Contrastive Learning}
    \STATE Aggregate future horizon: \(\mathbf{z} = \frac{1}{K} \sum_{k=1}^{K} \mathbf{e}^u_{t+k}\).
    \STATE Project current state: \(\mathbf{h}^{z} = \mathbf{W} \mathbf{h} + \mathbf{b}\).
    \STATE Calculate contrastive loss \(\mathcal{L}_{{FC}}\) using Equation (\ref{eq:loss_cl}).
    
    \STATE \textbf{// 4. Optimization}
    \STATE Total Loss: \(\mathcal{L} = \mathcal{L}_{{M}} + \mathcal{L}_{{FS}} + \lambda \mathcal{L}_{{FC}}\).
    \STATE Update parameters \(\Theta\) via backpropagation: \(\Theta \leftarrow \Theta - \eta \nabla_{\Theta} \mathcal{L}\).
\ENDFOR
\end{algorithmic}
\end{algorithm}

\subsection{Training and Inference}
Algorithm \ref{alg:training} details the training process of our method.
The final training objective comprises three components: the primary task loss $\mathcal{L}_{M}$, which calculates the prediction loss for the next item; the uncertainty-guided future supervision loss $\mathcal{L}_{FS}$ for the future $K-1$ steps beyond the next item; and the future-aware contrastive loss $\mathcal{L}_{FC}$, which serves as a regularization term. The total loss is formulated as:
\begin{equation}
    \mathcal{L} = \mathcal{L}_{{M}} + \mathcal{L}_{{FS}} + \lambda  \mathcal{L}_{{FC}},
\end{equation}
where $\lambda$ controls the weight of the contrastive loss. 

Algorithm \ref{alg:inference} details the inference process of our method. During the inference phase, only the primary task of next-item prediction is performed, while the two auxiliary future learning tasks are discarded, since they are utilized exclusively during the training phase to improve representation quality. Consequently, our framework enhances representation learning without introducing any additional computational overhead during inference.





\begin{algorithm}[t!]
\caption{Inference Procedure of the Proposed Framework}
\label{alg:inference}
\renewcommand{\baselinestretch}{1.2}\selectfont 
\begin{algorithmic}[1]
\REQUIRE Trained model parameters \(\Theta\) (including item embedding matrix \(\mathbf{M}\)), User's historical sequence \(\mathcal{S}^u_{1:t}\), Recommendation list size \(N\).
\ENSURE Top-\(N\) recommended items \(\mathcal{R}^u\) for the next timestep.

\STATE \textbf{// 1. Sequence Encoding}
\STATE Compute input embeddings \(\mathbf{H}\) corresponding to \(\mathcal{S}^u_{1:t}\).
\STATE Encode sequence: \(\mathbf{H}^{L} = \text{Transformer}(\mathbf{H})\).
\STATE Obtain current state representation: \(\mathbf{h} = \mathbf{H}^{L}[t]\).

\STATE \textbf{// 2. Score Computation and Ranking}
\STATE Compute prediction scores for all candidate items: \(\hat{\mathbf{y}} = \operatorname{softmax}(\mathbf{h} \mathbf{M}^\mathrm{T})\).
\STATE Rank all items in descending order based on their predicted probabilities in \(\hat{\mathbf{y}}\).
\STATE Select the Top-\(N\) items with the highest scores to form the recommendation list \(\mathcal{R}^u\).

\RETURN \(\mathcal{R}^u\)
\end{algorithmic}
\end{algorithm}

\section{Efficiency Analysis}
In this section, we analyze the time complexity of our proposed framework during both training and inference phases.

\subsection{{Training Efficiency}}
The computational cost of our framework during training comprises the sequential recommendation backbone and the two proposed auxiliary tasks.
\begin{itemize}
\item \textbf{Backbone Complexity:} Following the standard Transformer architecture \cite{vaswani2017attention}, the computational complexity is dominated by the self-attention mechanism. For a sequence of length $t$ and hidden dimension $d$, the complexity per layer is $\mathcal{O}(t^2 d + t d^2)$. With $L$ layers, the total backbone complexity is $\mathcal{O}(L(t^2 d + t d^2))$.
\item \textbf{Auxiliary Task Overhead:} 
The \textit{Uncertainty-Guided Future Supervision} introduces $K-1$ parallel projectors. As described in Eq. \ref{eq:projector}, each projector is implemented as a single linear layer followed by a ReLU activation. The computational complexity for generating $K-1$ future states is $\mathcal{O}((K-1) \cdot d^2)$, where $K$ is very small. However, thanks to the parallel design, these projections are implemented as a single batched matrix multiplication. Thus, the effective latency on parallel hardware (e.g., GPUs) is comparable to a single projection $\mathcal{O}(d^2)$, resulting in minimal overhead. The entropy calculation is linear with respect to the vocabulary size $\mathcal{O}(|\mathcal{V}|)$.
The \textit{Future-Aware Contrastive Learning} involves a pooling operation and a projection head. The pooling is $\mathcal{O}(K \cdot d)$. The projection head is a single linear layer, resulting in a complexity of $\mathcal{O}(d^2)$. 
\end{itemize}

\textbf{Comparison:} Although the proposed modules introduce a theoretical complexity of $\mathcal{O}(K d^2 + |\mathcal{V}|)$, the effective overhead is negligible. As established, $K$ is very small, and the parallel design reduces the latency of the projections to near $\mathcal{O}(d^2)$. This is insignificant compared to the standard sequential backbone, where the quadratic complexity of $\mathcal{O}(L(t^2 d + t d^2))$ dominates.

\subsection{{Inference Efficiency}}
A distinct advantage of our framework is its zero-overhead inference. As detailed in the previous section, the auxiliary components, specifically the uncertainty-guided future supervision and the future-aware contrastive learning, are strictly training-time regularization tools.
During the inference phase, these modules are detached. The model structure reverts exactly to the base sequence encoder (Equation \ref{rec_sequence_representation}) followed by the main prediction head. Therefore, the inference time complexity is identical to the underlying backbone model. Consequently, our method improves recommendation accuracy through enhanced representation learning without incurring any latency penalty in real-time deployment.

\section{Experiments}

In this section, we conduct extensive experiments to address the following research questions:

\begin{itemize}
\item \textbf{RQ1}: Does UFRec consistently outperform existing sequential recommendation baselines?
\item \textbf{RQ2}: How well does UFRec generalize across different backbone SR architectures?
\item \textbf{RQ3}: What is the contribution of each core component in UFRec, and is each component necessary for the overall performance?
\item \textbf{RQ4}: How do key hyperparameters affect the performance of UFRec?
\item \textbf{RQ5}: How robust is the proposed model against varying user sequence lengths and data sparsity?
\item \textbf{RQ6}: What is the role of uncertainty-guided modulation when utilizing future supervision signals?
\end{itemize}

\subsection{Experimental Settings}

\begin{table}[t] 
\renewcommand\arraystretch{1.1}
	\centering
	\caption{Dataset statistics.}  
 \setlength\tabcolsep{6.0pt}
 \scalebox{1.0}{
\begin{tabular}{lrrrrr}
\toprule
Datasets & \#Users  & \#Items  & \#Actions & Avg. Length & Density  \\
\midrule
Yelp &19,936 & 14,587 &207,952 &10.4 &0.07\%  \\
Sports &35,598&   18,357 & 296,337 &8.3  &0.05\%    \\
Beauty &22,363 &12,101  &198,502  &8.8  &0.07\%   \\
Office &4,905&   2,420 & 53,258 &10.9  &0.45\%    \\
\bottomrule
\end{tabular}}
\label{dataset}
\end{table}


\subsubsection{\textbf{Datasets}}
To comprehensively evaluate the effectiveness of our proposed approach, we conduct experiments on four real-world datasets from diverse application domains: Yelp\footnote{\url{https://www.yelp.com/dataset}} and three subsets from the Amazon review corpus\footnote{\url{http://jmcauley.ucsd.edu/data/amazon/}} (Sports, Beauty, and Office). Table \ref{dataset} summarizes the statistics of these collections. Specifically, the datasets are introduced as follows:

\begin{itemize}
    \item \textbf{Yelp}: This dataset serves as a standard and widely adopted benchmark for point-of-interest (POI) and local business recommendation. It provides comprehensive real-world data focusing on the interaction patterns between users and local businesses. The extensive user-item interaction graph makes it highly suitable for evaluating the effectiveness of recommendation models in capturing implicit user preferences for local services.
    
    \item \textbf{Sports (Sports and Outdoors)}: Derived from the extensive Amazon review corpus, this dataset focuses on the e-commerce domain of sporting goods, fitness equipment, and outdoor gear. It contains rich user-item interactions which capture users' diverse preferences in recreational activities.
    
    \item \textbf{Beauty}: Also extracted from the Amazon review corpus, this dataset represents user interactions with cosmetics, skincare, and personal care products. It exhibits distinct user purchasing behaviors and interaction patterns compared to other domains, providing a valuable scenario for evaluating the robustness and generalization capabilities of recommendation models across different types of consumer goods.
    
\item \textbf{Office}: As another subset of the Amazon corpus, this dataset records user interactions with various office-related items.
\end{itemize}

Consistent with established preprocessing protocols in recent recommendation literature \cite{liu2021contrastive,xie2022contrastive,qin2023meta}, we utilize the 5-core versions of these datasets. Specifically, the data has been reduced to extract subsets by filtering out users and items with fewer than five interaction records.

\subsubsection{\textbf{Evaluation Metrics}}
To evaluate model performance, we employ two widely recognized metrics: Hit Rate (HR) and Normalized Discounted Cumulative Gain (NDCG). Following established protocols \cite{zhang2024finerec,wang2024can,he2020lightgcn,kang2018self}, we report HR@\(m\) and NDCG@\(m\) for \(m \in \{10, 20\}\). 

For dataset partitioning, we utilize the standard chronological leave-one-out strategy. Specifically, for each user, the final interaction is designated for testing, the penultimate interaction for validation, and the remaining historical interactions for training. During the training process, we employed early stopping with a patience of 10 epochs based on the validation accuracy to prevent overfitting. Furthermore, to ensure an unbiased and fair evaluation, we evaluate the ranking of predictions across the entire item set, including users' historically interacted items, rather than relying on negative sampling.

\subsubsection{\textbf{Baseline Methods}}
To ensure a comprehensive assessment, we compare our method with 13 baseline methods, categorized into three groups: classical methods (GRU4Rec, Caser, SASRec, BERT4Rec), LLM-enhanced methods (LRD, LLM-ESR), and self-supervised learning-based methods ($\rm {{S}^{3}\text{-}{Rec}}$, CL4SRec, CoSeRec, ICLRec, DuoRec, ICSRec, FENRec).
\begin{itemize}
\item \textbf{GRU4Rec} \cite{hidasi2015session} applies recurrent neural networks (RNN) to sequential recommendation.
\item \textbf{Caser} \cite{tang2018personalized} utilizes a CNN-based approach to model high-order relationships in the context of sequential recommendation.
\item \textbf{SASRec} \cite{kang2018self} is the first work to utilize the self-attention mechanism for sequential recommendation.
\item \textbf{BERT4Rec} \cite{sun2019bert4rec} employs the BERT \cite{devlin2018bert} framework to capture the context information of user behaviors.
\item \textbf{LRD} \cite{yang2024sequential} is an LLM-based method. It leverages LLMs to discover new relations between items and reconstructs one item based on the relation and another item.
\item \textbf{LLM-ESR} \cite{liu2024llm} is also an LLM-based method. It addresses the long-tail problem by simultaneously leveraging collaborative signals and semantic information through the dual-view modeling and self-distillation.
\item $\rm {\textbf{S}^{3}\text{-}\textbf{Rec}}$ \cite{zhou2020s3} leverages four self-supervised objectives to uncover the inherent correlations within the data.
\item \textbf{CL4SRec} \cite{xie2022contrastive} proposes three random augmentation operators to generate positive samples for contrastive learning.
\item $\textbf{CoSeRec}$ \cite{liu2021contrastive} introduces two additional informative augmentation operators, building upon the foundation of CL4SRec. 
\item $\textbf{ICLRec}$ \cite{chen2022intent} clusters user interests into distinct categories and brings the representations of users with similar interests closer together.
\item \textbf{DuoRec} \cite{qiu2022contrastive} combines a model-level dropout augmentation and a sampling strategy for choosing hard positive samples.
\item \textbf{ICSRec} \cite{qin2024intent} is an improvement on ICLRec, further segmenting a user's sequential behaviors into multiple subsequences to generate finer-grained user intentions for contrastive learning.
\item \textbf{FENRec} \cite{huang2025future} enhances
sequential recommendation by employing time-dependent soft labeling to utilize future interactions and incorporating enduring hard negative samples.
\end{itemize}

\begin{table*}[t]
\renewcommand\arraystretch{1.15}
\centering
\caption{Performance comparison of different methods on four datasets (\textbf{HR@10 and NDCG@10}). Bold font indicates the best performance, while underlined values represent the second-best. Our method achieves state-of-the-art results among all methods, as confirmed by a paired t-test with a significance level of 0.01.}
\label{tab:compare_results_10}
\resizebox{\textwidth}{!}{%
\begin{tabular}{l|cc|cc|cc|cc}
\toprule
\multirow{2}{*}{\textbf{Model}} & \multicolumn{2}{c|}{\textbf{Yelp}} & \multicolumn{2}{c|}{\textbf{Sports}} & \multicolumn{2}{c|}{\textbf{Beauty}} & \multicolumn{2}{c}{\textbf{Office}} \\ \cmidrule(l){2-9} 
 & \textbf{HR@10} & \textbf{NDCG@10} & \textbf{HR@10} & \textbf{NDCG@10} & \textbf{HR@10} & \textbf{NDCG@10} & \textbf{HR@10} & \textbf{NDCG@10} \\ \midrule
GRU4Rec & 0.0364 & 0.0172 & 0.0178 & 0.0091 & 0.0325 & 0.0166 & 0.0543 & 0.0262 \\
Caser & 0.0389 & 0.0212 & 0.0261 & 0.0138 & 0.0341 & 0.0239 & 0.0581 & 0.0291 \\
SASRec & 0.0571 & 0.0306 & 0.0326 & 0.0184 & 0.0598 & 0.0308 & 0.0795 & 0.0345 \\
BERT4Rec & 0.0580 & 0.0316 & 0.0281 & 0.0142 & 0.0526 & 0.0282 & 0.0826 & 0.0374 \\ \midrule
LRD & 0.0698 & 0.0358 & 0.0397 & 0.0210 & 0.0684 & 0.0312 & 0.0926 & 0.0441 \\
LLM-ESR & 0.0672 & 0.0367 & 0.0426 & 0.0228 & 0.0752 & 0.0435 & 0.0975 & 0.0496 \\ \midrule
S3-Rec & 0.0608 & 0.0337 & 0.0336 & 0.0185 & 0.0619 & 0.0323 & 0.0937 & 0.0422 \\
CL4SRec & 0.0581 & 0.0312 & 0.0382 & 0.0217 & 0.0654 & 0.0350 & 0.0698 & 0.0325 \\
CoSeRec & 0.0608 & 0.0306 & 0.0426 & 0.0231 & 0.0726 & 0.0410 & 0.0792 & 0.0419 \\
ICLRec & 0.0596 & 0.0322 & 0.0429 & 0.0238 & 0.0741 & 0.0405 & 0.0928 & 0.0408 \\
DuoRec & 0.0702 & 0.0378 & 0.0471 & 0.0240 & 0.0830 & 0.0425 & 0.1028 & 0.0501 \\
ICSRec & 0.0711 & 0.0381 & 0.0482 & 0.0241 & 0.0839 & 0.0428 & {\ul 0.1045} & {\ul 0.0509} \\
FENRec & {\ul 0.0735} & {\ul 0.0394} & {\ul 0.0511} & {\ul 0.0244} & {\ul 0.0854} & {\ul 0.0435} & 0.1004 & 0.0502 \\ \midrule
\textbf{UFRec} & \textbf{0.0800} & \textbf{0.0426} & \textbf{0.0543} & \textbf{0.0257} & \textbf{0.0914} & \textbf{0.0459} & \textbf{0.1105} & \textbf{0.0535} \\ \midrule
Improv. & 8.84\% & 8.12\% & 6.26\% & 5.33\% & 7.03\% & 5.52\% & 5.74\% & 5.11\% \\ \bottomrule
\end{tabular}%
}
\end{table*}

\begin{table*}[t]
\renewcommand\arraystretch{1.15}
\centering
\caption{Performance comparison of different methods on four datasets (\textbf{HR@20 and NDCG@20}). Bold font indicates the best performance, while underlined values represent the second-best. Our method achieves state-of-the-art results among all methods, as confirmed by a paired t-test with a significance level of 0.01.}
\label{tab:compare_results_20}
\resizebox{\textwidth}{!}{%
\begin{tabular}{l|cc|cc|cc|cc}
\toprule
\multirow{2}{*}{\textbf{Model}} & \multicolumn{2}{c|}{\textbf{Yelp}} & \multicolumn{2}{c|}{\textbf{Sports}} & \multicolumn{2}{c|}{\textbf{Beauty}} & \multicolumn{2}{c}{\textbf{Office}} \\ \cmidrule(l){2-9} 
 & \textbf{HR@20} & \textbf{NDCG@20} & \textbf{HR@20} & \textbf{NDCG@20} & \textbf{HR@20} & \textbf{NDCG@20} & \textbf{HR@20} & \textbf{NDCG@20} \\ \midrule
GRU4Rec & 0.0637 & 0.0245 & 0.0279 & 0.0116 & 0.0499 & 0.0209 & 0.0961 & 0.0357 \\
Caser & 0.0659 & 0.0276 & 0.0352 & 0.0168 & 0.0510 & 0.0205 & 0.1042 & 0.0396 \\
SASRec & 0.0895 & 0.0392 & 0.0478 & 0.0222 & 0.0877 & 0.0378 & 0.1335 & 0.0478 \\
BERT4Rec & 0.0908 & 0.0393 & 0.0444 & 0.0185 & 0.0789 & 0.0349 & 0.1429 & 0.0515 \\ \midrule
LRD & 0.1085 & 0.0458 & 0.0605 & 0.0261 & 0.0980 & 0.0433 & 0.1491 & 0.0609 \\
LLM-ESR & 0.1064 & 0.0453 & 0.0646 & 0.0279 & 0.1068 & 0.0521 & 0.1526 & 0.0633 \\ \midrule
S3-Rec & 0.0958 & 0.0439 & 0.0510 & 0.0229 & 0.0937 & 0.0403 & 0.1562 & 0.0566 \\
CL4SRec & 0.0921 & 0.0392 & 0.0553 & 0.0244 & 0.0985 & 0.0410 & 0.1295 & 0.0492 \\
CoSeRec & 0.0974 & 0.0415 & 0.0636 & 0.0275 & 0.1031 & 0.0488 & 0.1368 & 0.0508 \\
ICLRec & 0.0965 & 0.0428 & 0.0639 & 0.0282 & 0.1059 & 0.0488 & 0.1524 & 0.0546 \\
DuoRec & 0.1079 & 0.0467 & 0.0692 & 0.0290 & 0.1202 & 0.0524 & 0.1619 & 0.0639 \\
ICSRec & 0.1086 & 0.0472 & 0.0712 & 0.0294 & 0.1208 & 0.0529 & {\ul 0.1645} & {\ul 0.0645} \\
FENRec & {\ul 0.1112} & {\ul 0.0489} & {\ul 0.0735} & {\ul 0.0300} & {\ul 0.1223} & {\ul 0.0536} & 0.1598 & 0.0632 \\ \midrule
\textbf{UFRec} & \textbf{0.1233} & \textbf{0.0535} & \textbf{0.0798} & \textbf{0.0321} & \textbf{0.1304} & \textbf{0.0555} & \textbf{0.1708} & \textbf{0.0680} \\ \midrule
Improv. & 10.88\% & 9.41\% & 8.57\% & 7.00\% & 6.62\% & 3.54\% & 3.82\% & 5.43\% \\ \bottomrule
\end{tabular}%
}
\end{table*}

\subsubsection{\textbf{Implementation Details}} 
We implemented all baseline methods using their officially released code. All experiments were conducted on a single NVIDIA A6000 GPU. For consistency, the embedding dimension was fixed at 64 across all models. Our sequential recommendation model utilizes a Transformer architecture consisting of two layers with two attention heads per layer. The model was trained using the Adam optimizer with a batch size of 256 and a learning rate of 0.001. Following standard practice in the literature \cite{qiu2022contrastive,qin2024intent}, the maximum sequence length was set to 50 for all datasets. We performed hyperparameter tuning on the validation set, selecting the coefficient \(\lambda\) from the set \(\{0.01, 0.05, 0.1, 0.2, 0.5, 1.0\}\). Similarly, \(K\) and \(\tau\) were tuned over the sets \(\{2, 3, 4, 5\}\) and \(\{1, 2, 3, 4, 5, 6\}\), respectively. We employed early stopping with a patience of 10 epochs based on the validation accuracy. All experiments were repeated three times to ensure reliability, and the results were averaged to provide a fair comparison.

\begin{figure*}[t]
  \centering
  \includegraphics[width=0.8\textwidth]{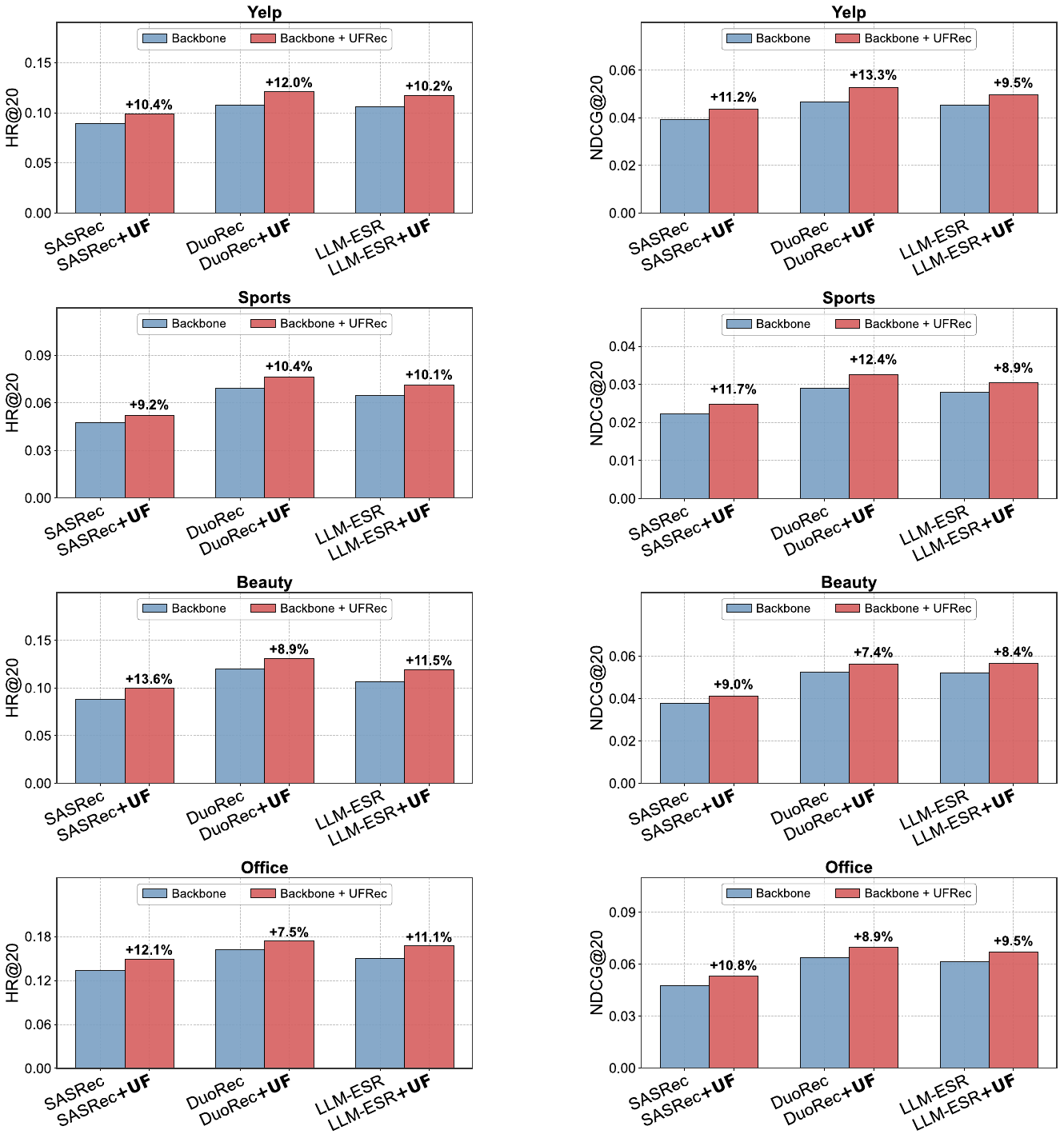}
  \caption{Generalizability analysis of UFRec across different sequential recommendation models. We compare the original performance of SASRec, LLM-ESR, and DuoRec against their variants enhanced with our framework (denoted as “+UF”).
  } 
   \label{diejia_experiments}
\end{figure*}

\subsection{Comparison Results (RQ1)}

Tables \ref{tab:compare_results_10} and \ref{tab:compare_results_20} present the comprehensive comparative results against 13 baseline methods across four datasets, evaluated by HR@$m$ and NDCG@$m$ (\(m \in \{10, 20\}\)). Each experiment is conducted three times, and the average values are reported. Overall, the performance of the evaluated methods exhibits distinct patterns based on their underlying paradigms, which can be summarized as follows:

\begin{itemize}
    \item \textbf{Performance of Traditional Methods:} Among the traditional approaches, GRU4Rec and Caser exhibit the weakest performance. While attention-based models like SASRec and BERT4Rec achieve noticeable improvements by capturing bidirectional or self-attentive item dependencies, they still demonstrate mediocre performance overall. This limitation primarily stems from their reliance on sparse interaction data without the guidance of auxiliary supervision signals.
    
    \item \textbf{Superiority of LLM-based Methods:} In contrast to traditional models, LLM-based methods (LRD, LLM-ESR) show highly competitive results. For instance, LLM-ESR significantly outperforms traditional baselines on the Beauty and Office datasets across all metrics. This substantial performance leap is attributed to the large language models' superior capacity for semantic understanding, which effectively alleviates the data sparsity issue inherent in pure ID-based sequential recommendation.
    
    \item \textbf{Effectiveness of Self-Supervised Learning (SSL):} SSL methods (S3-Rec, CL4SRec, CoSeRec, ICLRec, DuoRec, ICSRec, FENRec) consistently surpass traditional methods, as their auxiliary supervision signals enhance the utilization of limited data. Within this category, FENRec generally demonstrates superior performance compared to most other baselines on the Yelp, Sports, and Beauty datasets. This success is attributed to FENRec’s incorporation of future interactions as supervision signals via time-dependent soft labeling. However, on the Office dataset, ICSRec slightly edges out FENRec (e.g., achieving an HR@10 of 0.1045 compared to FENRec's 0.1004), indicating that FENRec's strategy of time-dependent soft labeling  lacks adaptability across varying data distributions.
    
    \item \textbf{State-of-the-Art Performance of UFRec:} Most importantly, our proposed method, UFRec, consistently and significantly outperforms all baseline methods across all four datasets and all evaluation metrics. For example, UFRec achieves remarkable relative improvements of up to 10.88\% in HR@20 and 9.41\% in NDCG@20 on the Yelp dataset. The fundamental limitation of the strongest baseline, FENRec, is that it applies future supervision indiscriminately to all samples, failing to account for the model's uncertainty regarding the current prediction. In contrast, UFRec quantifies uncertainty in the primary next-item prediction task using Shannon entropy, allowing it to dynamically adjust the weight of the auxiliary future supervision loss. This mechanism ensures that the model assigns less attention to future supervision when it lacks confidence in the immediate next step, thereby preventing noise propagation. These consistent improvements are validated by a paired t-test with a significance level of \(p < 0.01\), confirming the effectiveness and robustness of our uncertainty-guided approach.
\end{itemize}

\begin{table}[t]
\renewcommand\arraystretch{1.2}
\centering
\caption{Ablation study on all datasets.}
\label{tab:ablation}
\setlength\tabcolsep{5.8pt}
\scalebox{1.0}{
\begin{tabular}{l|l|cccc}
\toprule
 & Metric & {\makecell[c]{w/o $\mathcal{L}_{FS}$}}& {\makecell[c]{w/o UG}} & {\makecell[c]{w/o $\mathcal{L}_{FC}$}} & \textbf{UFRec}  \\
\midrule
\multirow{2}*{Yelp}  
&HR@20 &0.1202 &0.1187  &  0.1207 &  \textbf{0.1233}  \\
&NDCG@20 &0.0514 &  0.0511&  0.0516 & \textbf{0.0535}   \\
\midrule
\multirow{2}*{Sports}  
&HR@20 &0.0749 &0.0741  &0.0767  & \textbf{0.0798}  \\
&NDCG@20 &0.0303 &0.0295  &0.0307  &\textbf{0.0321}   \\
\midrule
\multirow{2}*{Beauty}  
&HR@20 &0.1268 &0.1232  &0.1281   & \textbf{0.1304}  \\
&NDCG@20 &0.0541 &0.0508  &0.0543    &\textbf{0.0555}  \\

\midrule
\multirow{2}*{Office}  
&HR@20   &0.1682   &0.1575   &0.1678   & \textbf{0.1708}  \\
&NDCG@20 &0.0661   &0.0620   &0.0653   &\textbf{0.0680}   \\
\bottomrule
\end{tabular}}
\label{ablation study}
\end{table}

\subsection{Generalizability to Different SR Models (RQ2)} \label{sec:general}
In this section, we evaluate the generalizability and flexibility of our proposed method, positioning it not merely as a standalone model, but as a versatile future learning framework capable of seamlessly integrating with and enhancing various mainstream sequential recommendation architectures. To comprehensively demonstrate this plug-and-play capability, we carefully selected representative models from three distinct paradigms to serve as base encoders: SASRec, representing the foundational self-attention mechanism in traditional methods; LLM-ESR, representing the cutting-edge semantic understanding capabilities of LLM-based approaches; and DuoRec, serving as a strong baseline for contrastive self-supervised learning methods. During the training phase, we preserved the original primary loss functions and network structures of these backbone models, while elegantly incorporating our uncertainty-guided future learning objectives, specifically the uncertainty-guided future supervision loss \(\mathcal{L}_{FS}\) and the future-aware contrastive loss \(\mathcal{L}_{FC}\), as auxiliary regularization signals. The comparative results, visually presented in Figure~\ref{diejia_experiments}, clearly indicate that the variants enhanced with our framework (denoted with the suffix ``+UF'') consistently and significantly outperform their original vanilla versions across all three model categories and across all evaluated datasets. Specifically, the integration of our framework yields remarkable relative performance gains, with improvements ranging from 7.5\% to 13.6\% in the HR@20 metric and 7.4\% to 13.3\% in the NDCG@20 metric. It is particularly noteworthy that even for highly optimized and advanced models like LLM-ESR and DuoRec, our method still manages to extract additional performance gains by effectively leveraging future interaction signals while dynamically mitigating uncertainty noise. Ultimately, these compelling findings validate that UFRec is a highly robust, model-agnostic framework capable of universally boosting the predictive accuracy of diverse sequential recommendation models.

\begin{figure}[t]
  \centering
  \includegraphics[width=0.78\columnwidth]{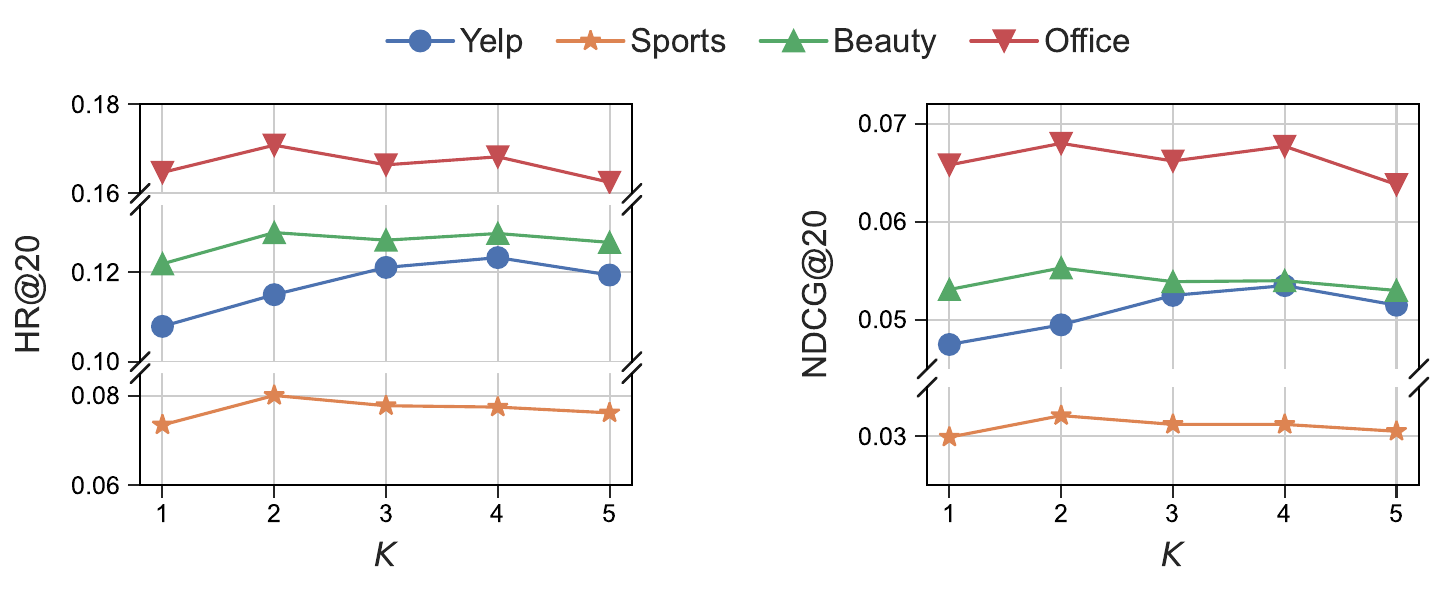}
  \caption{Hyperparameter study of maximum future
time step $K$ on four datasets.} 
   \label{param_study1}
\end{figure}

\subsection{Ablation Study (RQ3)}
In this section, we conduct an ablation study to evaluate the individual contribution of each core component within the UFRec framework. The results, summarized in Table \ref{tab:ablation}, demonstrate the impact of removing individual modules. We define the variants as follows: ``w/o $\mathcal{L}_{FS}$'' removes the uncertainty-guided future supervision entirely; ``w/o UG'' retains future supervision but removes the uncertainty guidance mechanism, meaning the supervision is no longer modulated by uncertainty; and ``w/o $\mathcal{L}_{FC}$'' removes the future-aware contrastive learning module. Overall, the results indicate that removing any component leads to performance degradation, confirming the necessity of each module. Specifically, the variants ``w/o $\mathcal{L}_{FS}$'' and ``w/o $\mathcal{L}_{FC}$'' exhibit significant performance drops, highlighting the critical roles of both auxiliary tasks. Notably, the ``w/o UG'' variant exhibits an even more severe decline than removing future supervision entirely. In this setting, future supervision is applied indiscriminately to all samples with equal weight, regardless of the model's prediction uncertainty. This drastic performance degradation validates the critical role of the uncertainty guidance mechanism in regulating future supervision.

\subsection{Hyperparameter Study (RQ4)}

\begin{figure}[t]
  \centering
  \includegraphics[width=0.78\columnwidth]{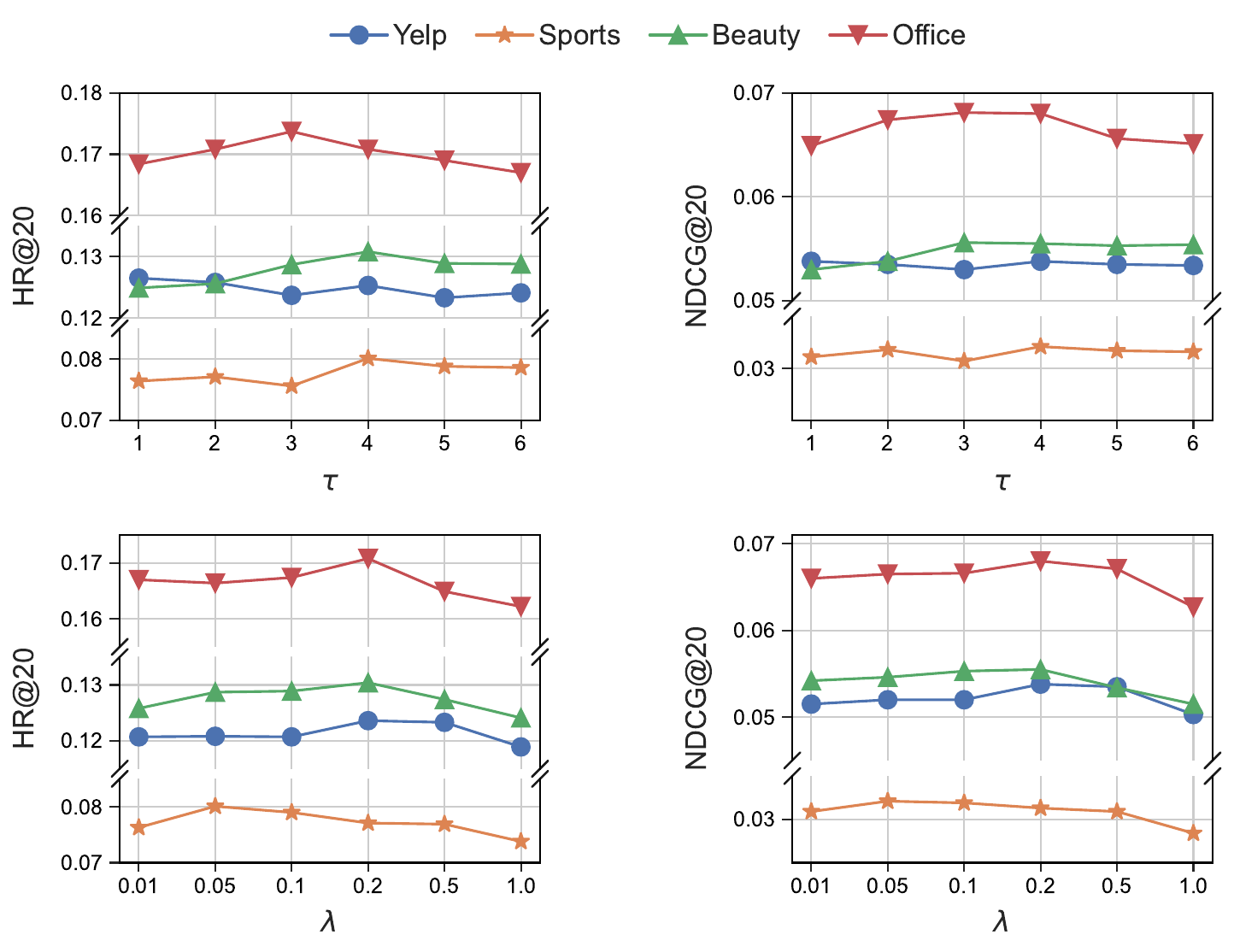}
  \caption{Hyperparameter study of $\tau$, and $\lambda$ on four datasets.} 
   \label{param_study2}
\end{figure}

In this section, we investigate the sensitivity and impact of three key hyperparameters essential to our framework: \(K\), \(\tau\), and \(\lambda\). Specifically, \(K\) represents the maximum future time step for look-ahead supervision, \(\tau\) denotes the temperature parameter controlling the uncertainty guidance in Equation (\ref{eq:weight}), and \(\lambda\) serves as the balancing weight for the future-aware contrastive loss. Figure~\ref{param_study1} illustrates the performance trends of HR@20 and NDCG@20 across the four datasets as \(K\) varies. We consistently observe that the recommendation accuracy initially improves as \(K\) increases, but subsequently declines after reaching a peak. Optimal performance is achieved at \(K=4\) for the Yelp dataset and at \(K=2\) for the other three datasets. We attribute this distinct trend to a fundamental trade-off between future data utilization and auxiliary task difficulty. While increasing \(K\) introduces additional supervisory signals from the future, thereby enhancing data efficiency and enriching the learned representations, larger values inherently increase the predictive difficulty of the auxiliary task. Consequently, excessively difficult auxiliary tasks can introduce noise and interfere with the primary next-item prediction objective, ultimately leading to performance degradation. Furthermore, the temperature parameter \(\tau\) critically controls the smoothing of the adaptive weight \(\omega\). As shown in Figure~\ref{param_study2}, the optimal value for \(\tau\) typically falls within the moderate range of 3 to 5. Values outside this optimal range result in a weight distribution that is either excessively sharp (acting like a rigid hard selection) or overly smooth (failing to distinguish between high and low confidence predictions), both of which lead to suboptimal performance. Finally, increasing the contrastive learning weight \(\lambda\) yields an initial improvement in overall model performance, followed by a noticeable decline. Empirical results indicate that the optimal values for \(\lambda\) generally fall within the relatively small range of 0.05 to 0.2. This trend is entirely anticipated, given that the contrastive learning loss primarily functions as a regularizer to refine representation learning; setting this weight too high causes the auxiliary contrastive loss to dominate the training process, thereby overshadowing and hindering the primary recommendation task.


\begin{table}[t]
\centering
\caption{Data statistics of the Yelp and Sports datasets across different user sequence length groups (i.e., \(=5\), \(6-8\), and \(>8\)).}
\label{tab:dataset_stats}
\renewcommand{\arraystretch}{1.1}
\begin{tabular}{l|c|c|c|c|c|c}
\hline
Datasets & \multicolumn{3}{c|}{Yelp} & \multicolumn{3}{c}{Sports} \\ \hline
\#length & =5 & 6-8 & >8 & =5 & 6-8 & >8 \\ \hline
\#users & 5235 & 7152 & 7549 & 11416 & 14209 & 9973 \\
\#items & 9968 & 12536 & 14539 & 15548 & 17477 & 17994 \\
\#actions & 26175 & 48284 & 133493 & 57080 & 95564 & 143693 \\
sparsity & 99.95\% & 99.95\% & 99.88\% & 99.97\% & 99.96\% & 99.92\% \\ \hline
\end{tabular}
\end{table}

\begin{figure}[t]
\setlength{\belowcaptionskip}{-1mm} 
  \centering
  \includegraphics[width=0.85\columnwidth]{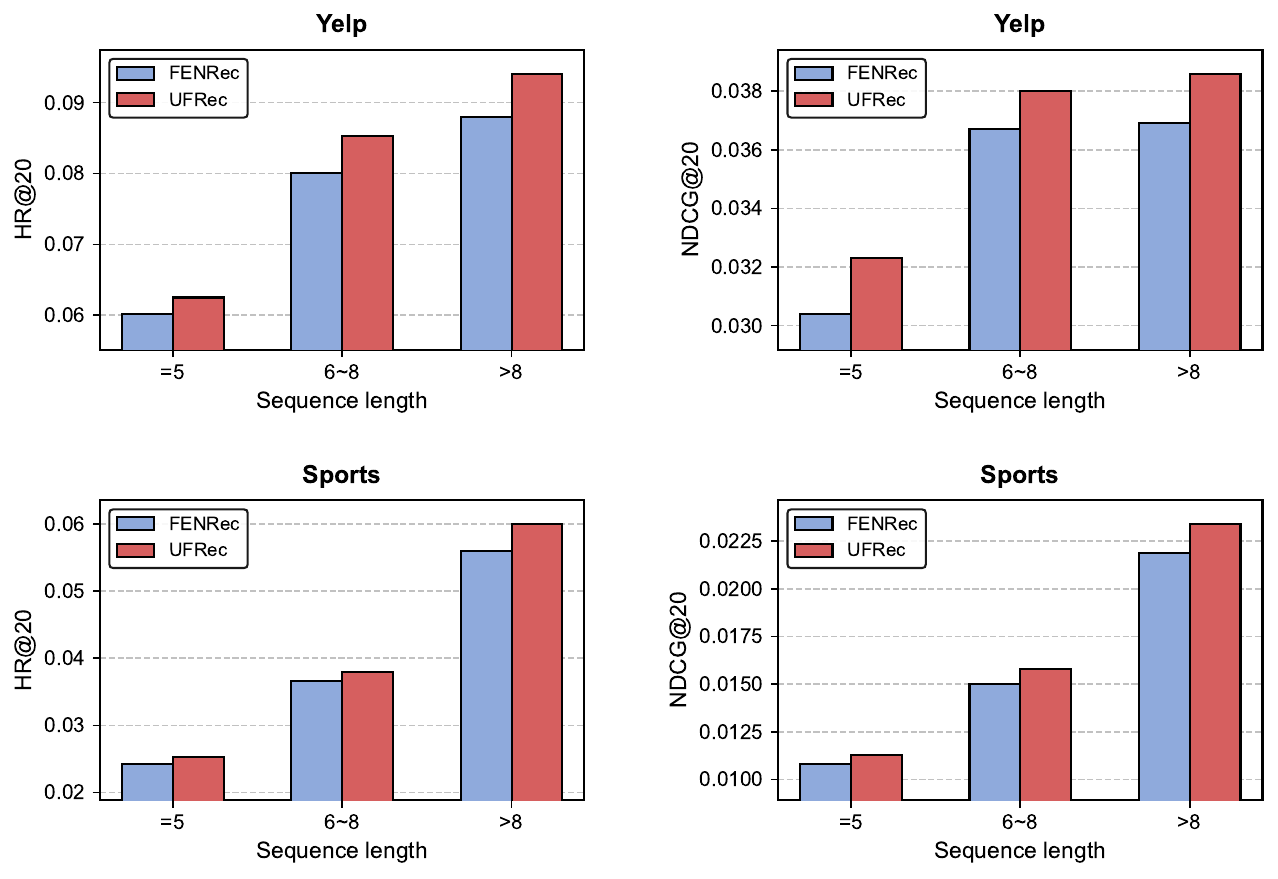}
  \caption{Performance comparison on different user groups.} 
   \label{group_study}
\end{figure}

\subsection{Robustness w.r.t. User Sequence Length (RQ5)}
To further examine the robustness of our model against varying degrees of data sparsity, we categorize user sequences into three distinct groups based on their historical length (i.e., \(=5\), \(6-8\), and \(>8\)). Specifically, the group with a length of \(=5\) represents extremely sparse users who possess the bare minimum of interactions, effectively simulating cold-start scenarios; the \(6-8\) group corresponds to users with moderate interaction histories; and the \(>8\) group denotes relatively active users equipped with richer historical contexts. The detailed statistical information regarding the users, items, actions, and sparsity levels for each group is summarized in Table~\ref{tab:dataset_stats}. Following this categorization, Figure~\ref{group_study} presents the performance comparison results on the Yelp and Sports datasets. By comparing our model with the strongest baseline FENRec, we make the following key observations: 1) The performance of both models inevitably deteriorates as interaction frequency decreases. This trend clearly highlights the profound influence of data sparsity, as shorter sequences provide insufficient historical context to accurately capture underlying user preferences. 2) Our method UFRec consistently outperforms the competitive baseline FENRec across every user group. Most notably, even for the group with the most limited data (sequence length of 5), our model maintains a significant performance lead. This demonstrates the positive impact of our uncertainty-aware future learning approach, which effectively extracts richer supervisory signals from limited interactions to alleviate data sparsity. Ultimately, these findings underscore the robustness and adaptability of our model across various degrees of data sparsity in user sequences.

\begin{figure}[t]
\setlength{\belowcaptionskip}{-1mm} 
  \centering
  \includegraphics[width=0.85\columnwidth]{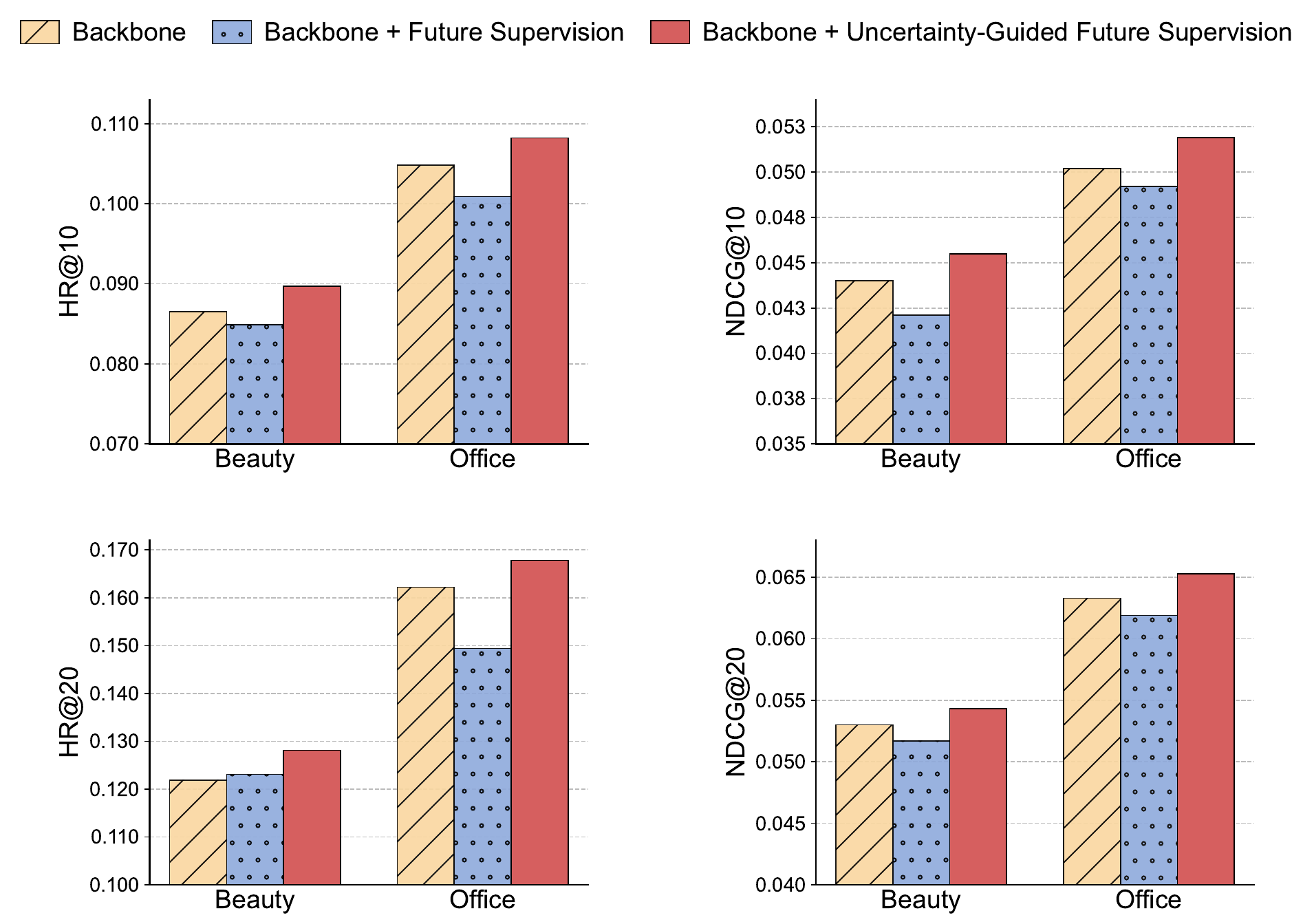}
  \caption{Impact of Uncertainty-Guided Modulation on Future Supervision. The results demonstrate that while indiscriminate future supervision (Backbone + FS) degrades performance, the uncertainty-guided approach yields significant improvements.} 
   \label{certainty_study}
\end{figure}

\subsection{In-depth Analysis of Uncertainty-Guided Modulation (RQ6)}
In this section, we provide further analysis to explicitly demonstrate the efficacy of uncertainty-guided modulation in enhancing future supervision. Specifically, employing the Transformer described in Section \ref{sec:backbone} as the foundational backbone, we compare the performance of the backbone integrated with standard, unweighted future supervision (Backbone + Future Supervision) against the backbone equipped with uncertainty-guided future supervision (Backbone + Uncertainty-Guided Future Supervision). The empirical results on the Beauty and Office datasets, illustrated in Figure \ref{certainty_study}, reveal that directly incorporating raw future supervision signals into the backbone not only fails to yield improvements but actually degrades overall performance. This suggests that the indiscriminate introduction of future supervision across all training instances likely introduces substantial noise into the learning process. This observation corroborates our core hypothesis: when the model exhibits high uncertainty in predicting the immediate next item—indicating an insufficient grasp of the user’s current dynamic state—forcing it to simultaneously predict distant future interactions overwhelms its learning capacity, turning otherwise valid future signals into a source of optimization interference. In contrast, integrating uncertainty-guided future supervision results in significant performance gains. This reaffirms that uncertainty-guided modulation acts as a crucial adaptive gate, effectively regulating the intensity of future supervision signals based on the model's real-time uncertainty regarding the primary task. Consequently, the model dynamically acquires additional supervision signals when confident in its current predictions, while intelligently attenuating the influence of future supervision to focus on the primary task when uncertain, thereby safeguarding the representation learning process and enhancing overall performance.

\section{Conclusion}
In this paper, we identified a critical limitation in existing sequential recommendation models: their predominant reliance on immediate next-item supervision, which restricts their ability to capture long-term preference evolution. While incorporating multi-step future interactions offers a promising solution, we demonstrated that applying such supervision indiscriminately can introduce detrimental noise. To overcome this, we proposed {UFRec}, a model-agnostic framework designed to adaptively leverage future interactions. 
At the core of UFRec are two complementary modules. First, the {Uncertainty-Guided Future Supervision} mechanism quantifies the model's confidence of the primary next-item prediction using Shannon entropy, dynamically modulating the weight of multi-step future supervisions. Second, the {Future-Aware Contrastive Learning} module treats the future trajectory as a holistic entity, explicitly ensuring that a user's current representation is more similar to their own future horizon than to the future trajectories of other users.
Extensive experiments conducted on four real-world benchmark datasets validate the superiority of our approach. UFRec consistently outperforms 13 state-of-the-art baselines. Furthermore, our in-depth analyses demonstrate the framework's strong generalizability across different backbone architectures and its robustness against severe data sparsity. As a training-time regularization strategy, UFRec enhances representation learning without introducing any additional computational overhead during the inference phase, making it highly efficient for real-world deployment. 

In future work, we plan to explore more sophisticated methods for modeling future horizons, such as leveraging Large Language Models to infer latent future intents, and extending our uncertainty-guided mechanism to other sequence-based tasks.

\begin{acks}
This work is supported by the Early Career Scheme (No.CityU 21219323) and the General Research Fund (No.CityU 11220324) of the University Grants Committee (UGC), the NSFC Young Scientists Fund (No.9240127), and the Donation for Research Projects (No.9229164 and No.9229216).
\end{acks}

\bibliographystyle{ACM-Reference-Format}
\bibliography{sample-base}

\end{document}